\documentstyle[12pt]{article}

\newlength{\dinwidth}
\newlength{\dinmargin}
\begin{document}
\titlepage
\begin{flushright}
DTP/97/58 \\
July 1997 \\
\end{flushright}

\begin{center}
\vspace*{2cm}
{\Large \bf Deep inelastic events containing two \\ 
forward jets at HERA} \\
\vspace*{1cm}
J.\ Kwiecinski$^1$, C.A.M.\ Lewis$^2$ and A.D.\ Martin$^2$
\end{center}

\vspace*{0.5cm}
\begin{tabbing}
$^1$xxxx \= \kill
\indent $^1$ \> Department of Theoretical Physics, H.\
Niewodniczanski Institute of Nuclear Physics, \\
\indent $\quad$ ul.\ Radzikowskiego 152, 31-342 Krakow, Poland.
\\

\indent $^2$ \> Department of Physics, University of Durham,
Durham, DH1 3LE, UK. \\

\end{tabbing}

\vspace*{2cm}

\begin{abstract}
We use the BFKL equation to calculate the rate of deep inelastic
scattering events containing two forward jets (adjacent to the
proton remnants) at HERA.  We compare the production of two
forward jets with that of only one forward jet (the \lq \lq
Mueller" process).  We obtain a stable prediction for this two to
one jet ratio, which may serve as a measure of the BFKL vertex
function.
\end{abstract}

\newpage

\noindent{\large \bf I.~Introduction} 

One of the most striking discoveries at the electron-proton
collider, HERA, at DESY, is the steep rise of the proton
structure function $F_2 (x, Q^2)$ with decreasing Bjorken $x
\equiv Q^2 / 2 p \cdot q$, where $Q^2 \equiv -q^2$.  Here $p$ is
the four-momentum of the proton and $q$ is the four-momentum
transfer of the electron probe.  The behaviour of the structure
function at small $x$ is driven by the gluon through the process
$g \rightarrow q \overline{q}$.  The behaviour of the gluon
distribution at small $x$ is itself predicted from perturbative
QCD via the Balitskij-Fadin-Kuraev-Lipatov (BFKL) equation, which
resums the leading $[\alpha_S \ln (1/x)]^n$ contributions.  This
predicts a characteristic $x^{-\lambda}$ singular behaviour in
the small $x$ regime which results from the summation of soft
multi-gluon emissions, which effectively correspond to the sum of
gluon ladders (Fig.~1) where the transverse momenta $k_{Ti}$ of
the emitted gluons are unordered along the chain.  For fixed
$\alpha_S$ the BFKL exponent $\lambda = \overline{\alpha}_S 4 \ln
2$, where $\overline{\alpha}_S \equiv 3 \alpha_S/\pi$.  It is
this increase in the gluon distribution with decreasing $x$ which
produces the corresponding rise of the structure function.  Of
course there are uncertainties due to subleading corrections,
which will reduce the value of the exponent $\lambda$, and from
the treatment of the infrared $k_T$ region in the numerical
solution of the BFKL equation.  A recent description of $F_2$
along these lines, but also including the effects of
Dokshitzer-Gribov-Lipatov-Altarelli-Parisi (DGLAP) evolution
\cite{DGLAP}, can be found in ref. \cite{STASTO}.

However, the situation is more complicated.  DGLAP evolution,
which resums the $\alpha_S \ln Q^2$ contribution can on its own
produce an increase in $F_2$ with decreasing $x$, and with a
suitable choice of input distributions at scale $Q_0^2$ can give
an excellent description of $F_2$ at small $x$.  This makes it
difficult to distinguish whether the steep rise observed for the
structure function has a major component from $\ln (1/x)$
effects, or is simply due to $\ln Q^2$ evolution from a non-
perturbative input.  The observable $F_2$ is just too inclusive
to act as a discriminator of the underlying small $x$ dynamics.

In 1990 Mueller \cite{M} proposed that the less inclusive
quantity of DIS plus an identified energetic forward jet may
serve as a useful probe of small $x$ dynamics.  The idea is to
isolate DIS $(x, Q^2)$ events containing a jet $(x_j, k_{Tj}^2)$
with $x \ll x_j$ and $k_{Tj}^2 \sim Q^2$, see Fig.~1.  Here
$k_{Tj}$ and
$x_{j}p$ are the transverse and longitudinal momentum components
of the jet.  We can thus expect a $(x/x_j)^{- \lambda}$ behaviour
in a kinematic region free of DGLAP evolution\footnote{The
$[\alpha_S \ln Q^2]^n$ contributions which are resummed by DGLAP
come from the region in which the transverse momenta of the
emitted gluons are strongly ordered along the chain, that is $Q^2
\gg k_{Tn}^2 \gg \ldots k_{Tj}^2$.  DGLAP evolution is thus
neutralized by the choice $Q^2 \sim k_{Tj}^2$.}. Moreover, for
large $x_j \sim 0.1$, the parton distributions of the proton are
well known and so the Mueller proposal removes the uncertainty
arising from the introduction of a non-perturbative input.  That
is we study DIS off a known parton, rather than off the proton. 
Recently data for the process have become available and the
comparison with BFKL predictions is encouraging \cite{H1}.

The observed DIS + forward jet data sample contains a fraction
of events in which two jets are identified, each with $p_T$
greater than the experimental cut.  Our concern here is to extend
the BFKL formalism to study and to estimate the rate for this
process.  We also present the predictions as the ratio of 1 to 2
forward jet production.

The structure of the paper is as follows.  In Sec.~II we use the
data for DIS + one forward jet to determine the normalisation of
our numerical solution of the BFKL equation.  In Sec.~III we
discuss the extension of the BFKL formalism to include an extra
identified gluon jet, and then in Sec.~IV we present results for
the cross section of DIS + two identified forward jets.  Our
conclusions are given in Sec.~V. \\

\noindent{\large \bf II.~DIS + one forward jet}

The differential cross section for deep inelastic scattering $(x,
Q^2)$ containing a forward jet $(x_j, k_{Tj}^2)$ is of the form
\begin{equation}
\label{eq:a1}
\frac{\partial \sigma}{\partial x \partial Q^2 \partial x_j
\partial k_{Tj}^2} \; = \; \frac{4 \pi \alpha^2}{x Q^4} \: \left[
(1 - y) \frac{\partial F_2}{\partial x_j \: \partial k_{Tj}^2} \:
+ \: {\textstyle \frac{1}{2}} y^2 \: \frac{\partial
F_T}{\partial x_j \partial k_{Tj}^2} \right]
\end{equation}
where $y = Q^2/sx$ with $\sqrt{s}$ equal to the centre-of-mass
energy of the electron-proton collision.  The differential
structure functions
\begin{equation}
\label{eq:a2}
-x_j \: \frac{\partial F_i}{\partial x_j \partial k_{Tj}^2} \; =
\; \frac{3 \alpha_S (k_{Tj}^2)}{\pi k_{Tj}^4} \: \left(\sum_a x_j
f_a (x_j, k_{Tj}^2) \right) \: \Phi_i (x/x_j, k_{Tj}^2, Q^2)
\end{equation}
where the $\Phi_i$ represent the photon-gluon process shown by
the blob in Fig.~2(a).  The factor $k_{Tj}^4$ arises from the
gluon propagators.  We are interested in events with small
$x/x_j$ so $x_j$ should be taken as large as is experimentally
feasible.  Eq.~(2) assumes strong ordering of the longitudinal
momentum at the gluon-parton $a$ vertex, so that $x_j$ of the
exchanged parton, which occurs in the parton density $f_a (x_j,
k_{Tj}^2)$, is to a good approximation that of the outgoing jet. 
Assuming $t$ channel pole dominance we have 
\begin{equation}
\label{eq:a3}
\sum_a \: f_a \; = \; g + \textstyle{\frac{4}{9}} \: (q +
\overline{q})
\end{equation}
where $g$, $q$ and $\overline{q}$ are the gluon, quark and
antiquark densities respectively. To include the running of
$\alpha_S$ it is convenient to work in terms of the functions
\cite{KMS}
\begin{equation}
\label{eq:a4}
H_i (z, k_t^2) \; = \;  \overline{\alpha}_S (k_T^2) \Phi_i (z,
k_T^2, Q^2)
\end{equation}
with $i = 2, T$.  These functions satisfy the BFKL equations
\begin{eqnarray}
\label{eq:a5}
H_i (z, k_T^2, Q^2) & = & H_i^{(0)} (z, k_T^2, Q^2) \nonumber \\
\\
& + & \overline{\alpha}_S (k_T^2) \: \int_z^1 \: \frac{d
z^{\prime}}{z^{\prime}} \: \int \: \frac{d^2 q_T}{\pi q_T^2} \:
\left[ \frac{k_T^2}{k_T^{\prime 2}} \: H_i (z^{\prime},
k_T^{\prime 2}, Q^2) \: - \: H_i (z^{\prime}, k_T^2, Q^2) \Theta
(k_T^2 - q_T^2) \right] \nonumber \\ \nonumber
\end{eqnarray}
where $k_T^{\prime 2} \: = \: (\mbox{\boldmath$k$}_T +
\mbox{\boldmath$q$}_T)^2$.  Note that the $q_T^2 \rightarrow 0$
singularity is removed by the cancellation between the real and
virtual contributions.  Due to the strong ordering at the $g a a$
vertex the transverse momentum of the BFKL gluon is that of
the forward jet, that is $k_T^2 \simeq k_{Tj}^2$.  In eq.(5) we
impose a cut-off $k_T^{\prime 2} \; (k_T^2) > k_0^2$ (with $k_0^2
\sim 1 {\rm GeV}^2)$ in order to avoid the Landau pole
singularity of $\alpha_S (k_T^2)$.  The driving
terms $H_i^{(0)} = \overline{\alpha}_S (k_T^2) \Phi_i^{(0)}$ are
given by the quark box (plus the crossed
box) contributions to the $\gamma g$ fusion process
\begin{eqnarray}
\label{eq:a6}
\Phi_T^{(0)} & = & \sum_q e_q^2 \: \frac{Q^2}{4 \pi^2} \:
\alpha_S (k_T^2) \int_0^1 d \beta \int d^2 \kappa^{\prime}
\nonumber \\
&& \left\{[\beta^2 + (1 - \beta)^2]\left(
\frac{{\mbox{\boldmath$\kappa$}}}{D_1}
\: - \: \frac{{\mbox{\boldmath$\kappa$}} -
{\mbox{\boldmath$k$}}_T}{D_2} \right)^2 \: + \: m_q^2 \left(
\frac{1}{D_1} -
\frac{1}{D_2} \right)^2 \right\} \\
\Phi_L^{(0)} & = & \sum_q e_q^2 \: \frac{Q^4}{\pi^2} \:
\alpha_S(k_T^2) \int_0^1 d \beta \int d^2 \kappa^{\prime} \beta^2
(1 - \beta)^2 \: \left( \frac{1}{D_1} - \frac{1}{D_2} \right)^2
\nonumber \\ \nonumber
\end{eqnarray}
where $\kappa^{\prime} = \kappa - (1 - \beta) k_T$ and
\begin{eqnarray}
\label{eq:a7}
D_1 & = & \kappa^2 + \beta (1 - \beta) Q^2 + m_q^2 \nonumber \\
\\
D_2 & = & (\mbox{\boldmath$\kappa$} -
\mbox{\boldmath$k$}_T)^2 + \beta (1 - \beta) Q^2 + m_q^2.
\nonumber \\ \nonumber
\end{eqnarray}
For small $z$, eq.(5) generates the characteristic $z^{-
\lambda}$ behaviour with $\lambda = \overline{\alpha}_S \: 4
\ln 2$ for fixed coupling $\alpha_S$.  DIS + forward jet events
have been discussed within the BFKL formalism with
fixed coupling \cite{FIX}.  

Our objective is to calculate the ratio of DIS
events containing
two, as compared to one, forward jets. Since the contributions
from the input $\Phi_i^{(0)}$ largely cancel, for simplicity we
do not include charm in the equations for the quark box and we
assume massless quarks.  In the $z^{\prime}$ integration in
(\ref{eq:a5}) we impose a cut $z < z_0 \neq 1$, such that $x_j >
x$.  This determines the point, $z_0$, at which the evolution in
$\ln (x_j/x)$ starts, that is the point at which the gluon chain
is matched onto the quark box $\Phi_i^{(0)}$.  Essentially, we
take $z_0$ as a free parameter which is chosen to normalise the
prediction for the DIS + one forward jet cross section (as a
function of $x$) to the measured values.  We find $z_0 \simeq
0.5$.  If we were to enhance the driving terms by 20\%, say, to
allow for a charm contribution, then the starting input value
$z_0$ would be about 0.1.

To make the comparison between the prediction and the data we
must impose the experimental kinematic
cuts used in the measurement of the DIS + jet events. We adopt
the cuts used by the H1 collaboration \cite{H1}.  That is the
forward jet is required to lie in the region
\begin{equation}
\label{eq:a8}
7^{\circ} < \theta_j < 20^{\circ}, \quad E_j > 28.7{\rm GeV},
\quad k_{Tj} > 3.5{\rm GeV}
\end{equation}
whereas the outgoing electron is constrained to the domain
\begin{equation}
\label{eq:a9}
160^{\circ} < \theta_e < 173^{\circ}, \quad E_e > 11{\rm GeV},
\quad y > 0.1
\end{equation}
in the HERA frame.  Finally H1 require that
\begin{equation}
\label{eq:a10}
\textstyle{\frac{1}{2}} Q^2 < k_{jT}^2 < 2 Q^2
\end{equation}
so as to minimize the DGLAP $\ln Q^2$ effects.

We use these cuts to calculate the DIS + forward jet cross
section and compare with the H1 data in Fig.~3.  We find that the
choice $z_0 = 0.5$ leads to a good description of the data.  It
is relevant for our study of the 2 forward jet rate to note the
effect of the upper phase space cut on the variable $y$.  The
solid and dashed histograms in Fig.~3 correspond to using $y
<0.5$ and $y < 1$ respectively.  We see that the predictions at
small $x$ are sensitive to this cut.  In fact if the next
smallest $x$ bin, $\Delta x = (0.0001, 0.0005)$ had been shown we
would have seen a turnover due to the depletion of events caused
by the lack of phase space. \\

\noindent{\large \bf III. DIS + two forward jets}

Here we extend the BFKL formalism so as to be able to predict the
rate for the production of an extra jet in the forward direction.
The process
\begin{equation}
\label{eq:a11}
\gamma^* + p \rightarrow j_1 + j_2 + X
\end{equation}
is shown in Fig.~2(b).  We require that both jets are resolved
and have transverse momentum greater than some minimum cut,
$k_{Tji}^2 > \mu^2$ with $i = 1,2$.  The differential cross
section for this process is given by
\begin{eqnarray}
\label{eq:a12}
&& \frac{\partial \sigma}{\partial x \partial Q^2 \: \partial
x_{j1}
\: \partial k_{Tj1}^2 \: \partial x_{j2} \: \partial
k_{Tj2}^2}
\; = \; \nonumber \\
\\
&& \indent \indent \indent \indent \indent \frac{4 \pi
\alpha^2}{x Q^4}
\: \left[ (1 - y)
\frac{\partial
F_2}{\partial x_{j1} \: \partial k_{Tj1}^2 \: \partial x_{j2} \:
\partial k_{Tj2}^2} \: + \: {\textstyle \frac{1}{2}} y^2 \:
\frac{\partial F_T}{\partial x_{j1} \: \partial k_{Tj1}^2 \:
\partial x_{j2} \: \partial k_{Tj2}^2} \right], \nonumber \\
\nonumber
\end{eqnarray}
where the differential structure functions
\begin{eqnarray}
\label{eq:a13}
&& -x_{j2} \frac{\partial F_i}{\partial x_{j1} \: \partial
k_{Tj1}^2
\: \partial x_{j2} \: \partial k_{Tj2}^2}  = \; \nonumber \\
\\
&& \indent \indent \indent \indent \indent \frac{1}{2 \pi} \:
\int_0^{2
\pi} d \phi \; \Phi_i (x/x_{j2},
k_u^2, Q^2) \; \frac{\overline{\alpha}_S (k_{Tj2}^2)
k_{Tj1}^2}{k_{Tj2}^2 \: k_u^2} \: \frac{\overline{\alpha}_S
(k_{Tj1}^2)}{k_{Tj1}^4} \: \sum_a \: f_a (x_{j1}, k_{Tj1}^2),
\nonumber \\ \nonumber
\end{eqnarray}
where the variables are shown in Fig.~2(b).  The variable $\phi$
is the azimuthal angle between the two forward jets.  Because
$$
k_u^2 \; = \; ({\mbox{\boldmath$k$}}_{Tj1} +
{\mbox{\boldmath$k$}}_{Tj2})^2
$$
the integral over $\phi$ is non-trivial.  The general structure
of formula (\ref{eq:a13}) for the production of two forward jets
is readily understood as an extension of the
single jet formula, (\ref{eq:a2}).  The BFKL functions $\Phi_i$
describe the gluon radiation in the upper part of the gluon
emission chain, which is shown as a blob in the Fig.~2(b).  These
functions are obtained by solving the BFKL equation exactly as
described in Sec.~II, and are indeed normalised so as to describe
the DIS + one forward jet data.  Since both of the resolved jets
are required to be in the forward region $x_{j2} \lower .7ex\hbox{$\;\stackrel{\textstyle
<}{\sim}\;$}
x_{j1} \sim O (1)$, we do not enter the strongly ordered
configuration $x_{j2} \ll x_{j1} \sim O (1)$ and so we can
neglect the effect of soft (BFKL) gluon radiation emitted in the
rapidity interval between the two jets.  The remaining noteworthy
feature of (\ref{eq:a13}) is the presence of the BFKL vertex
function, $k_{Tj1}^2 / k_{Tj2}^2 \: k_u^2$ controlling the
emission of jet 2. 

When the second jet is produced in the central region (that is
when $x_{j2} \ll x_{j1})$ then the gluon radiation in the
rapidity interval between the two jets can no longer be
neglected.  We would need to use the formalism developed in
ref.\ \cite{DEC}. \\

\noindent{\large \bf IV.~Predictions of the DIS + two forward jet
rate}

We calculate the cross section for DIS + two forward jets from
eqs.(\ref{eq:a12}) and (\ref{eq:a13}), using the kinematic cuts
listed in eqs.(\ref{eq:a8}-\ref{eq:a10}).  We present the cross
section in Fig.~4.  To gain insight into the effect of the cuts
we also include the
results obtained using the enlarged domain
\begin{equation}
\label{eq:a14}
{\textstyle \frac{1}{2}} Q^2 < k_{Tj}^2 < 5 Q^2
\end{equation}
for each of the two jets.  For the experimental identification of
two jets we require a minimum angular separation between
the jets.  We therefore impose a cut on the $(\eta, \phi)$ space
of the two partons producing the forward jets.  For illustration
we show the results for 
\begin{equation}
\label{eq:a15}
\sqrt{(\Delta\eta)^2 \: + \: (\Delta \phi)^2} \: > R
\end{equation}
for four different values of $R$, namely $R = 0, 1, 1.7$ and 2,
where $\Delta \eta$ and $\Delta \phi$ are the differences in the
pseudorapidities and azimuthal angles of the two parton jets.

From Fig.~4 we see that the two jet cross section has a turnover
for the smallest $x$ bin for the lowest set of curves , which
correspond to the cut $k_{Tj}^2 / Q^2 < 2$.  The effect is not
apparent when we relax the constraint to $k_{Tj}^2 / Q^2 < 5$. 
Also we note that the cross section is appreciably larger for the
higher kinematic cut, $k_{Tj}^2 / Q^2 < 5$.  First we quantify
the increase.  When we change the $k_{Tj}^2 / Q^2$ cut from 2 to
5 the cross section for DIS + single forward jet increases by a
factor of about 2.5 for the smallest $x$ bin that is shown,
$\Delta x \: = \: (0.0005, \: 0.001)$, but increases only by a
factor of about 1.5 for the largest $x$ bin, $\Delta x \: = \:
(0.003, \: 0.0035)$.  Similarly, the cross section for DIS + two
forward jets increases by factors of 5.5 and 2.5 for these two
$\Delta x$ bins respectively.  The explanation for the large
increase of cross section as we relax the upper limit on
$k_{Tj}^2$ from $2 Q^2$ to $5 Q^2$ can be found by inspecting
Fig.~5.

Fig.~5 shows the differential cross section for DIS + two forward
jets, for the $x$ bin $\Delta x \: = \: (0.001, \: 0.0015)$, as a
function of the transverse momentum of the jets, which are taken
to have equal magnitude $k_{Tj1}^2 = k_{Tj2}^2$.  We take jet
resolution cut, $k_{Tj}^2 > \mu^2$, to be $\mu = 3.5 {\rm GeV}$. 
The curves on Fig.~5 correspond to the differential cross section
for different values of $Q^2$.  When we make the cut $k_{Tj}^2 /
Q^2 < 2$, we remove a substantial contribution to the cross
section from the low $Q^2$ region.  The effect is enhanced by the
jet resolution cut $k_{Tj}^2 > \mu^2$, which requires the cross
section to be zero when $\mu^2 / Q^2 < 2$, as indicated by the
vertical arrows on Fig.~5, which occur above the {\it lower}
bound $k_{Tj}^2 / Q^2 = \frac{1}{2}$ for both $Q^2 = 15$ and $12
{\rm GeV}^2$ if we take $\mu = 3.5 {\rm GeV}$.  Thus the
resolution constraint has significant impact at the lower values
of $Q^2$.  The $x$ dependence of the cross section for different
kinematic constraints is shown in Table 1.

The turnover that is apparent at small $x$ in the lower set of
histograms in Fig.~4 is simply due to the lack of phase space. 
The reason the cross section predictions for the smaller $x$ bins
are particularly sensitive to phase space restrictions, is that
such cuts kill the important low $Q^2$ contributions, see Fig.~5.

\begin{table}
\begin{center}
\begin{tabular}[b]{|l||c|c|c|c|} \hline
& & & & \\
\raisebox{1.5ex}[0pt]{$\quad \quad \quad x {\rm bin}$} &
\raisebox{1.5ex}[0pt]{$\displaystyle{\frac{k_{Tj}^2}{Q^2}} < 2$}
&
\raisebox{1.5ex}[0pt]{$\displaystyle{\frac{k_{Tj}^2}{Q^2}} < 3$}
&
\raisebox{1.5ex}[0pt]{$\displaystyle{\frac{k_{Tj}^2}{Q^2}} < 4$}
&
\raisebox{1.5ex}[0pt]{$\displaystyle{\frac{k_{Tj}^2}{Q^2}} < 5$}
\\
\hline
$0.0001 - 0.0005$ & $\; 1.9$ & $\; \: 7.5$ & 15.8 & 25.5
\\
$0.0005 - 0.001$ & $\; 7.2$ & 17.8 & 29.0 & 39.6 \\
$0.001 \; - \: 0.0015$ & $\; 6.0$ & 12.8 & 19.4 & 25.2 \\
$0.0015 - 0.002$ & $\; 4.4$ & $\; 8.6$ & 12.3 & 15.3 \\
$0.002 \; - \: 0.0025$ & $\; 3.2$ & $\; 5.9$ & $\; \: 8.0$
& $\; \: 9.7$ \\
$0.0025 - 0.003$ & $\; 2.3$ & $\; 4.1$ & $\; \: 5.4$ & $\; \:
6.4$
\\
$0.003 \; - \: 0.0035$ & $\; 1.7$ & $\; 2.8$ & $\; \: 3.7$
&
$\;
\: 4.3$
\\
\hline
\end{tabular}
\end{center}
\caption{The dependence of the DIS + two forward jet cross
section (in $pb$) on the kinematic constraint $\frac{1}{2} Q^2 <
k_{Tj}^2 < n Q^2$ for various values of $n$.  We show the
dependence for different intervals of $x$.  We take $R = 0, \; y
< 1$ and $\mu = 3.5 {\rm GeV}$.}
\end{table}

The BFKL functions $\Phi_i$ are common to the calculation of both
the single and two forward jet rates, see eqs.(\ref{eq:a2}) and
(\ref{eq:a13}).  Thus the ambiguities in the calculation should
be reduced in the prediction of the ratio $\sigma_{2 {\rm jet}} /
\sigma_{1 {\rm jet}}$.  The predictions are shown in Fig.~6, and
also Fig.~7 which shows the dependence of the ratio on the jet
resolution parameter $\mu$ (but using larger $\Delta  x$ bins). 
As anticipated from the previous discussion we see that the ratio
increases significantly when we relax the cut from $k_{Tj}^2 /
Q^2 < 2$ to $k_{Tj}^2 / Q^2 <5$.  We see that the ratio shows an
increase as we move from lower to higher values of $x$. The
reason is that the ordering $x_{j2} < x_{j1}$ of the two emitted
jets means that the reduction of the cross section due to
decreasing evolution length is less for 2 forward jets than for
single jet production.  The ratio is insensitive to the choice of
the upper limit for $y$.  This is illustrated in Table 2 for two
different jet separation cuts.

We conclude that the ratio is an observable which can give
information on the underlying small $x$ dynamics, and in
particular can act as a measure of the BFKL vertex function which
occurs in (\ref{eq:a13}).  The first experimental estimates of
the ratio have been made \cite{DER}, but so far only to give an
indication of the size of the effect.  A value of about 4\% is
quoted for the ratio of 2-jet/1-jet events for a slightly
different choice of kinematic variables than those of
(\ref{eq:a8})-(\ref{eq:a10}) above, namely for the domain
$$
6^{\circ} < \theta_j < 20^{\circ}, \quad x_j > 0.025, \quad
k_{Tj} > 5 {\rm GeV},
$$
$$
160^{\circ} < \theta_e < 173^{\circ}, \quad E_e > 12 {\rm GeV},
\quad 0.1 < y < 1, 
$$
$$
0.0002 < x < 0.002, \quad \frac{1}{2} \: Q^2 < k_{jT}^2 < 4 Q^2.
$$
For this kinematic region we predict a 2-jet/1-jet ratio of 6.6\%
at the parton level.

\begin{table}
\begin{center}
\begin{tabular}[b]{|c|c|c|c|c||c|c|c|c|} \hline
& \multicolumn{4}{|c||}{$R = 1.7$} & \multicolumn{4}{|c|}{$R =
2$}
\\ \cline{2-9}
$\mu$ & \multicolumn{2}{|c|}{$k_{Tj}^2 / Q^2 < 2$} &
\multicolumn{2}{|c||}{$k_{Tj}^2 / Q^2 < 5$} &
\multicolumn{2}{|c|}{$k_{Tj}^2 / Q^2 < 2$} &
\multicolumn{2}{|c|}{$k_{Tj}^2 / Q^2 < 5$} \\ \cline{2-9}
(GeV) & $y < 0.5$ & $y < 1$ & $y < 0.5$ & $y < 1$ & $y <0.5$ &
$y
< 1$
& $y < 0.5$ & $y < 1$ \\ \hline
3.5 & 3.2 & 3.5 & 7.1 & 7.1 & 2.8 & 3.1 & 6.3 & 6.4 \\
5 & 3.2 & 3.6 & 6.7 & 6.9 & 2.9 & 3.2 & 6.1 & 6.2 \\
6 & 3.1 & 3.4 & 6.3 & 6.4 & 2.7 & 3.1 & 5.7 & 5.8 \\ \hline
\end{tabular}
\end{center}
Table 2: The dependence of the 2:1 forward jet ratio on
the kinematical $(k_{Tj}^2 / Q^2)$ and phase space $(y)$ cuts. 
The cross section is shown for the $x$ bin $\Delta x = 0.0005 -
0.002$. $\mu$ is the jet resolution parameter, $k_{Tj}^2 >
\mu^2$, and $R$ the jet separation parameter of
eq.(\ref{eq:a15}).
\end{table}

A characteristic feature of the effects of the BFKL chain is the
angular decorrelation of the jets \cite{AGKM}.  Due to the lack
of $k_T$-ordering the $k_u$ gluon of Fig.~2(b) can bring
significant transverse momentum into the two jet system and hence
considerably broaden the \lq \lq back-to-back" peak in the
azimuthal distribution $\Delta \phi \; = \; \phi_1 - \phi_2$. 
Sample distributions are shown in Fig.~8.

In order to remove the infrared $k_T^2 \rightarrow 0$ infinities
in the BFKL equation we impose the cut $k_T^2 =
(\mbox{\boldmath$k$}_{Tj1} + \mbox{\boldmath$k$}_{Tj2})^2 > 1
{\rm GeV}^2$.  This, together with the ambiguity due to
hadronization, means that the distribution near the back-to-back
configuration cannot be predicted.  Rather it is the tails of the
distribution
which will characterize the lack of $k_T$ ordering. 
Unfortunately the jet separation cut, which experimentally is
likely to be $R \sim 1.7$, will effectively remove the whole of
the tail of the distribution, see Fig.~8.  The azimuthal
decorrelation is thus unlikely to be a way of identifying the
underlying small $x$ dynamics. \\

\noindent{\large \bf V.~Conclusions}

The measurement of DIS scattering events $(x, Q^2)$ containing a
very forward energetic jet $(x_j, k_{Tj}^2)$, with $x_j \gg x$
and $k_{Tj}^2 \sim Q^2$, has long been advocated as a favourable
way of investigating the dynamics which underlie small $x$
physics.  The forward region is defined by cuts of the type given
in eq.(\ref{eq:a8}) and (\ref{eq:a9}), and jet events are
collected which typically lie in the region
\begin{equation}
\label{eq:a16}
\textstyle{\frac{1}{2}} Q^2 < k_{Tj}^2 < 2 Q^2
\end{equation}
subject also to the jet resolution cut, $k_{Tj}^2 > \mu^2$.  To
obtain sufficient statistics the first measurements take $\mu =
3.5 {\rm GeV}$, but ideally as the integrated luminosity at HERA
improves, it will be preferable to obtain a DIS data sample
restricted to jets with higher transverse momentum. It is already
known that the DIS +
forward jet data sample contains a small fraction of events with two
identified jets \cite{DER}.  The main purpose of this paper is to
predict the fraction, and to study the properties, of these
events.  Our results are shown by
the lower set of histograms in Fig.~6 for $\mu = 3.5 {\rm GeV}$,
and in Fig.~7(a) for three different values of $\mu$.  We see
that the fraction of 2 jet events for $R = 1.7$ is about 2.5\%
for the $x$ interval $(0.5 - 1) \times 10^{-3}$ rising to about
5\% for $x = (2.5 - 3) \times 10^{-3}$, see Fig.~6. 
The fraction is rather insensitive (i) to the value of the jet
resolution parameter $\mu$, see Fig.~7(a), and (ii) to
ambiguities in the function $\Phi_i$ describing the BFKL gluon
chain since it is common to both the 1 jet and 2 jet predictions
and tends to cancel in the ratio.  The experimental confirmation
of the predicted 2 jet fractions will therefore serve as a check
on the BFKL vertex function which occurs in (\ref{eq:a13}).  

The domain (\ref{eq:a16}) is chosen such that $k_{Tj}^2 \sim Q^2$
so as to suppress DGLAP gluon emission.  We also presented
results for
the larger domain $\frac{1}{2} Q^2 < k_{Tj}^2 < 5 Q^2$ simply to
gain insight into the behaviour of the DIS + two forward jet
cross section. \\

\noindent{\large \bf Acknowledgements}

We thank Albert De Roeck and Ewelina Mroczko for useful
discussions.  J.K. thanks Grey College and the Department of Physics
at the University of Durham for their warm hospitality.  C.A.M.L
thanks the Particle Physics and Astronomy
Research Council for financial support.  This work has
been supported in part by Polish State Committee for Scientific
Research Grant No. 2 P03B 089 13, and by the EU
under Contracts Nos. CHRX-CT92-0004 and CHRX-CT93-357. \\

\newpage
\noindent{\large \bf Figure Captions}
\begin{itemize}
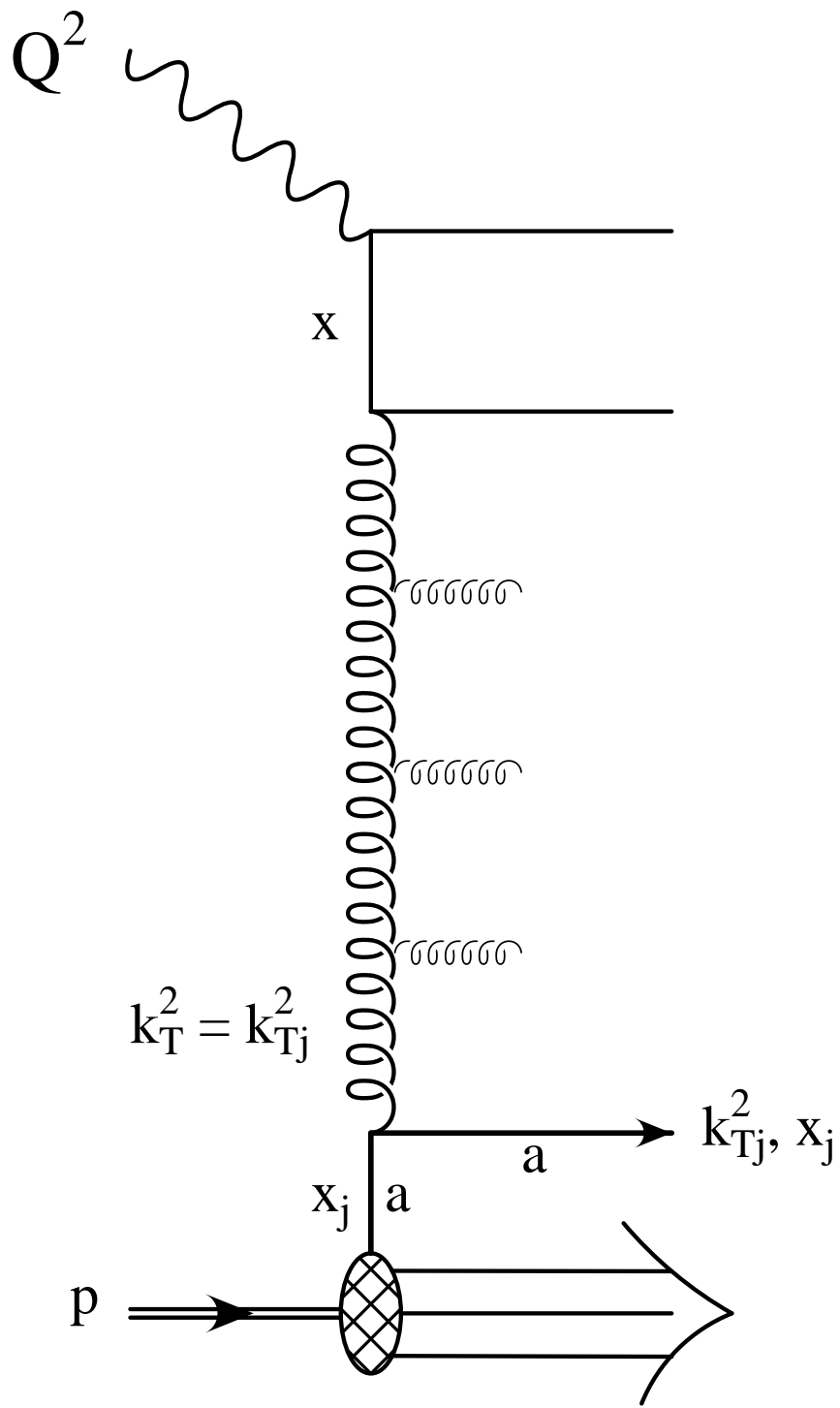
\item[Fig.~1] Diagrammatic representation of a deep inelastic
event containing an identified forward jet with longitudinal and
transverse momentum components $x_j p$ and $k_{Tj}$ respectively.
The photon scatters from the gluon chain (via the quark \lq box')
described by the BFKL equation which resums the (soft) gluon
emissions.  Parton $a$ can be either a gluon or a quark.

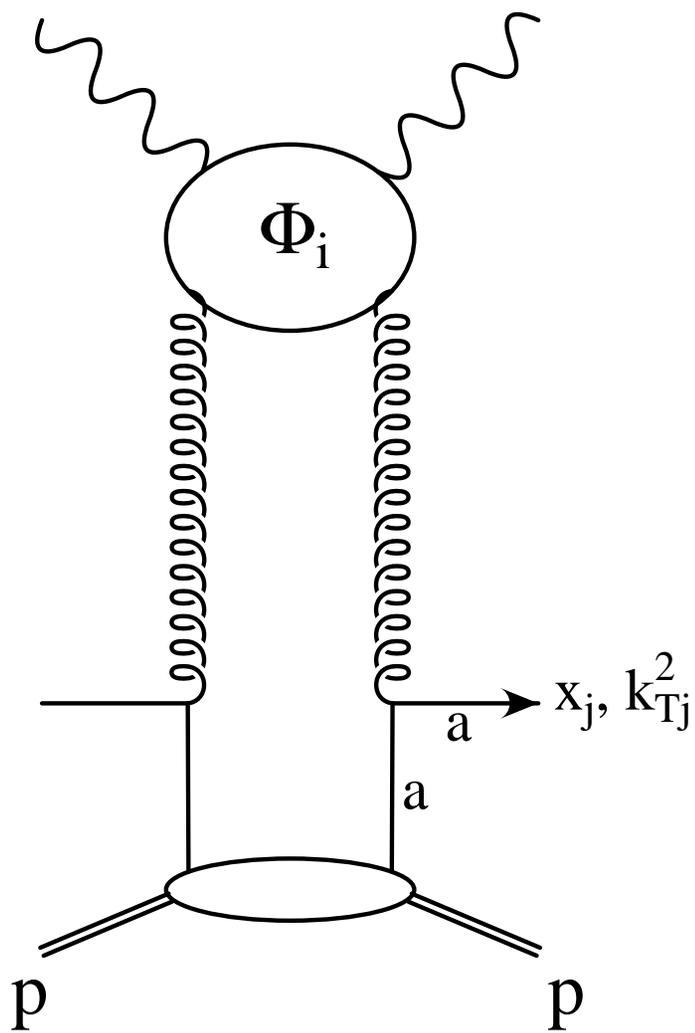
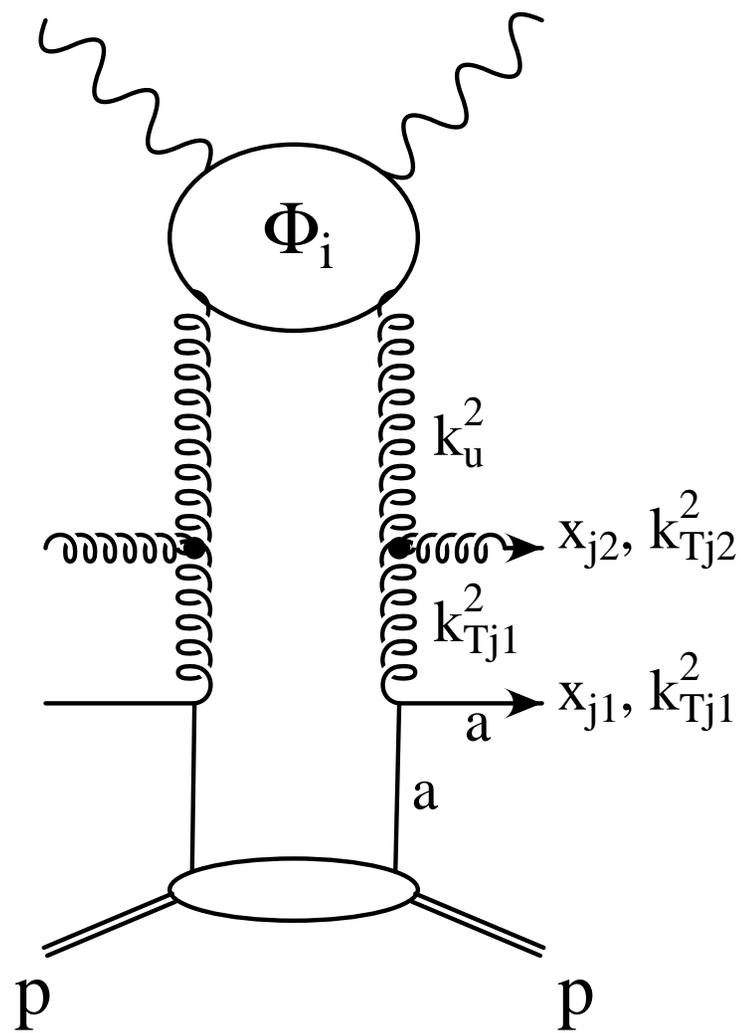
\item[Fig.~2] The diagrams giving the cross section for (a) DIS +
one forward jet and (b) DIS + two forward jets.  In each case the
upper blob represents the BFKL functions $\Phi_i$ (which in the
absence of the BFKL gluon emissions would, to lowest order, be
simply given by the quark box and crossed box, $\Phi_i^{(0)})$. 
The dot represents the BFKL vertex function \lq \lq measured" by
the 2 jet/1 jet ratio.

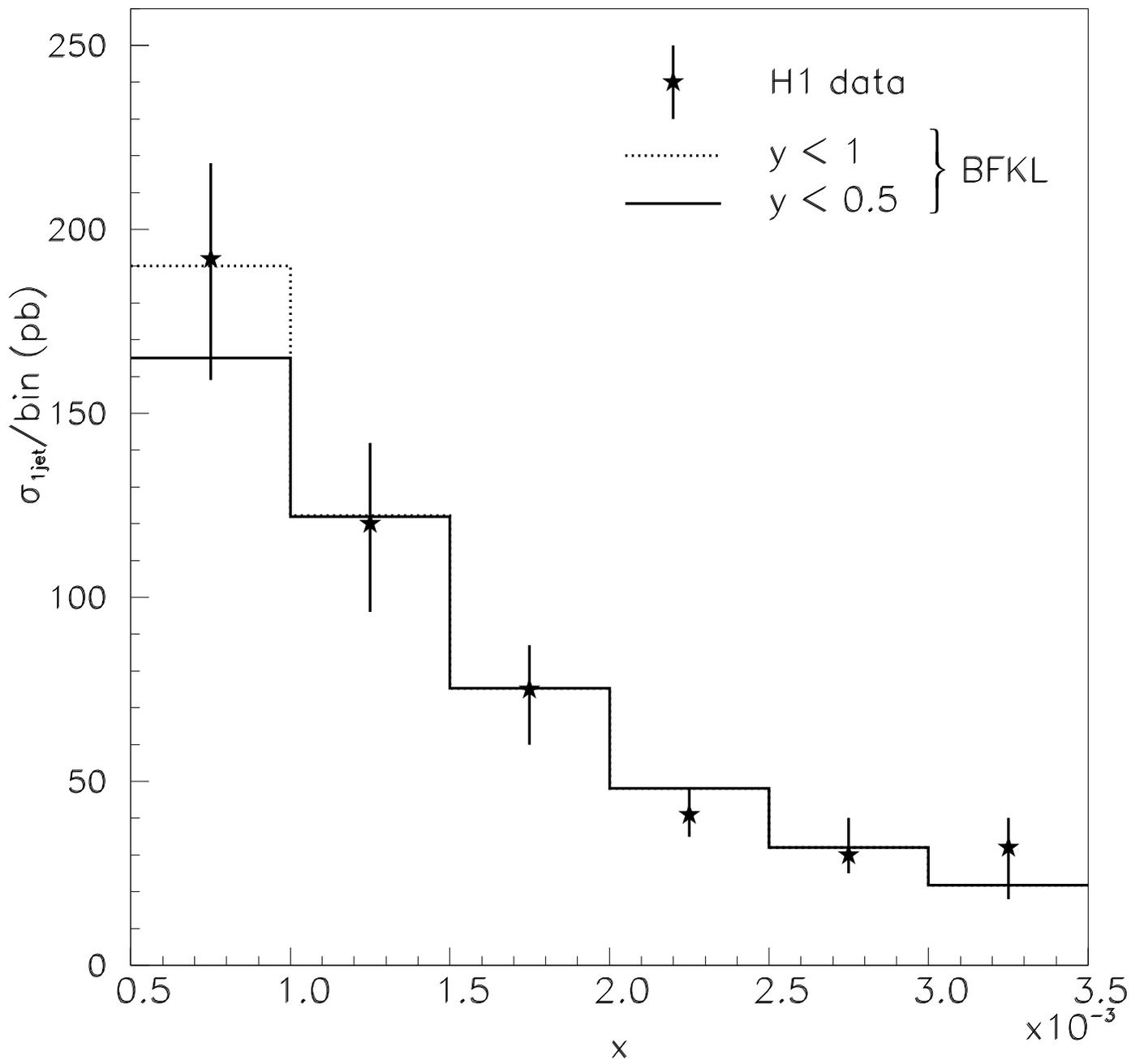
\item[Fig.~3] The DIS + forward jet cross section compared with
recent H1 data \cite{H1}. The phase space cuts are given in
eqs.(\ref{eq:a8}-\ref{eq:a10}). The resummed BFKL kernel is
normalized by the choice $z_0 = 0.5$.  The solid (dashed)
histograms correspond to $y < 0.5$ and $y < 1$ respectively.

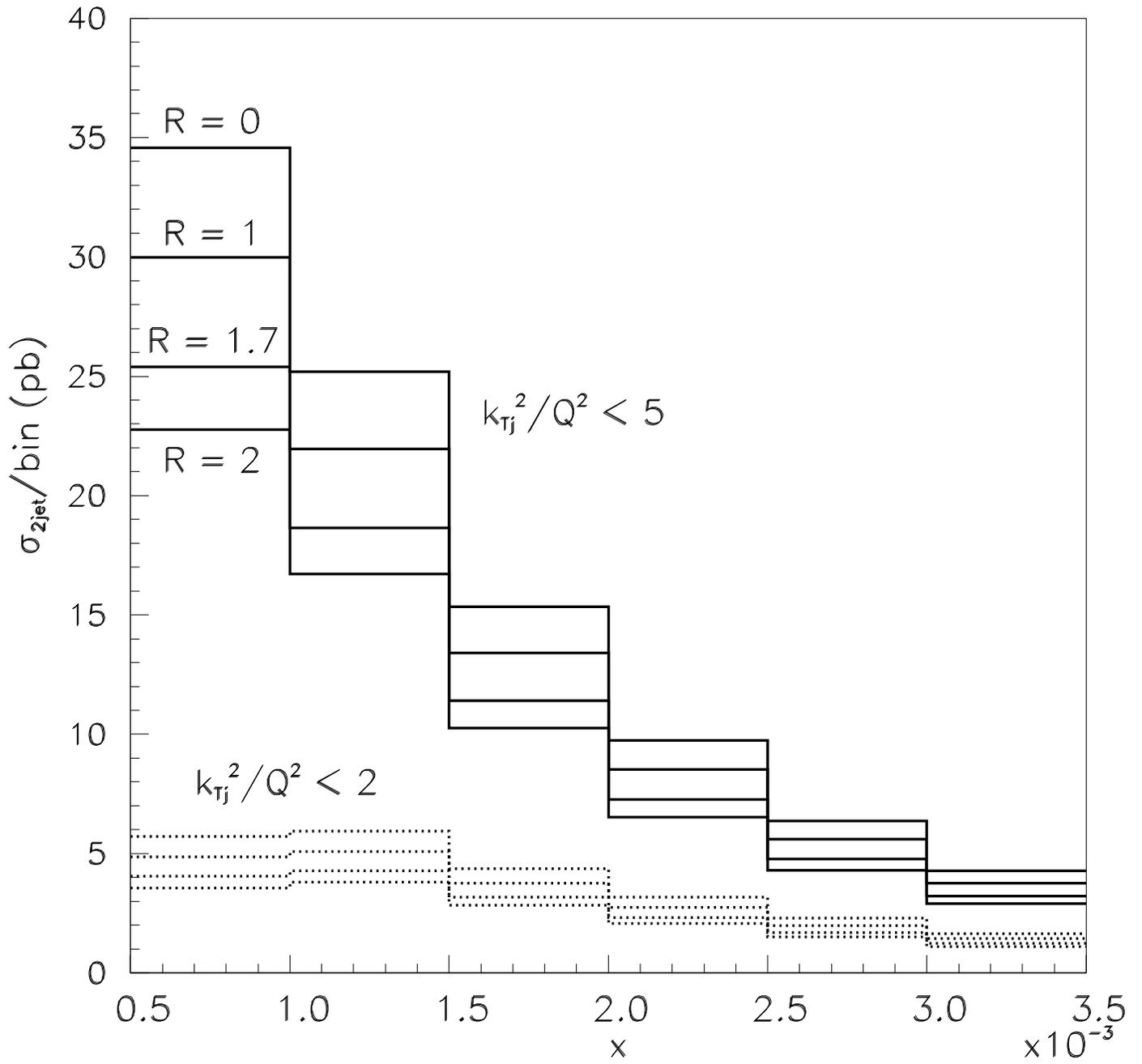
\item[Fig.~4] The DIS + two forward jet cross section $\sigma_2$
in the same $x$ bins as for the DIS + forward jet results.  We
take $y < 0.5$. $\sigma_2$ is plotted for $k_{Tj} > 3.5 {\rm
GeV}$ for two different choices of the upper $k_{Tj}$ cut:
$k_{Tj}^2 / Q^2 < 2$ and 5.  The results are shown for four
different cuts on the minimum separation between the two forward
jets: $R = 0, 1, 1.7$ and 2.  The smaller the separation cut, the
higher the cross section.

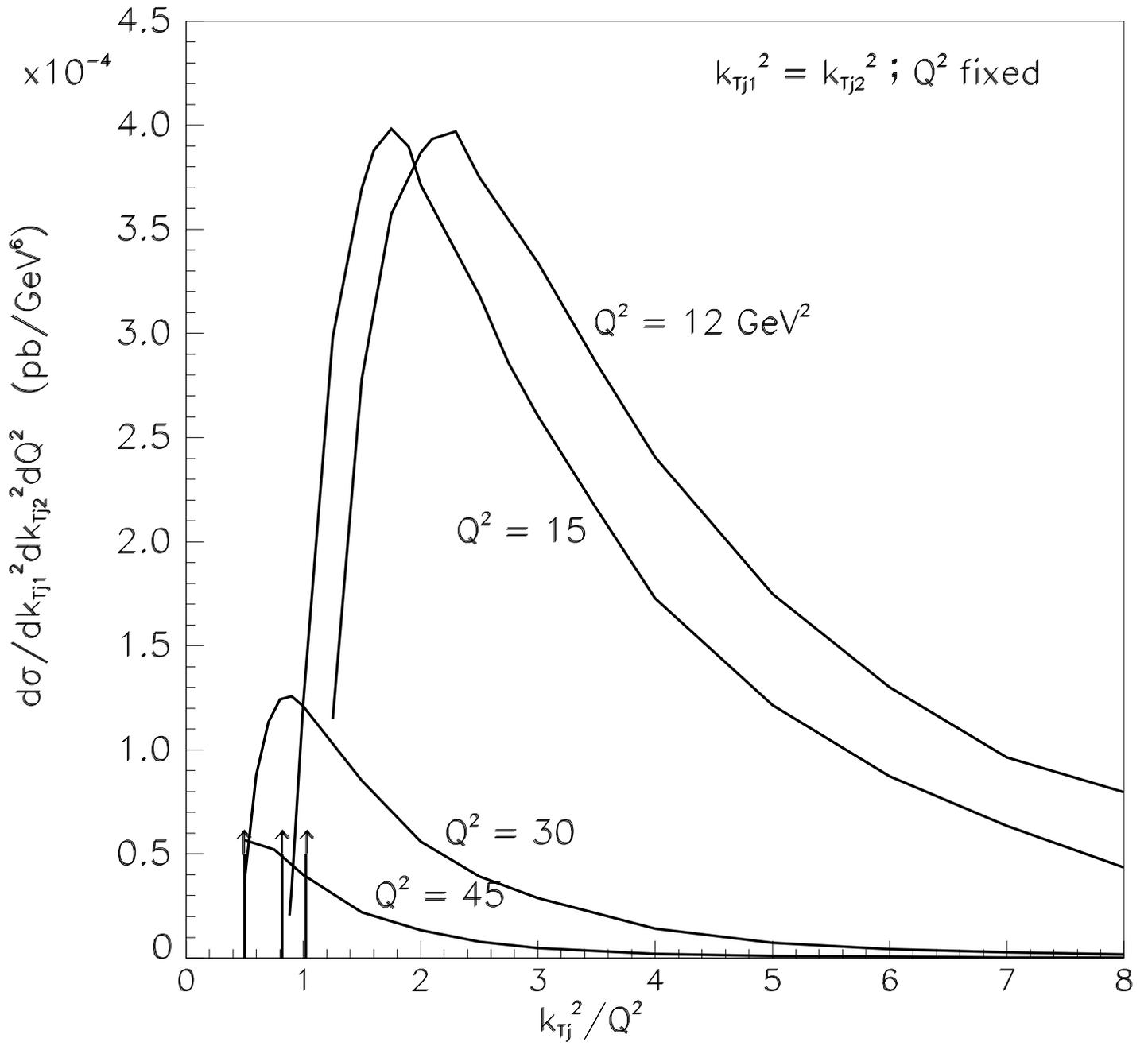
\item[Fig.~5] The differential cross section for DIS + two
forward jets as a function of $k_{Tj}^2 / Q^2$, for different
values of $Q^2$.  We take $k_{Tj1}^2 = k_{Tj2}^2, \; k_{Tj}^2 /
Q^2 > \frac{1}{2}, \; R = 0$ and the jet resolution cut off $\mu
= 3.5 {\rm GeV}$.  The cross section (\ref{eq:a12}) is integrated
over $x_{j1}$ and $x_{j2}$ and is shown for the $x$ bin $\Delta x
= (0.001, \;
0.0015)$.  The lower cut-offs are indicated by vertical arrows
for the different values of $Q^2$.

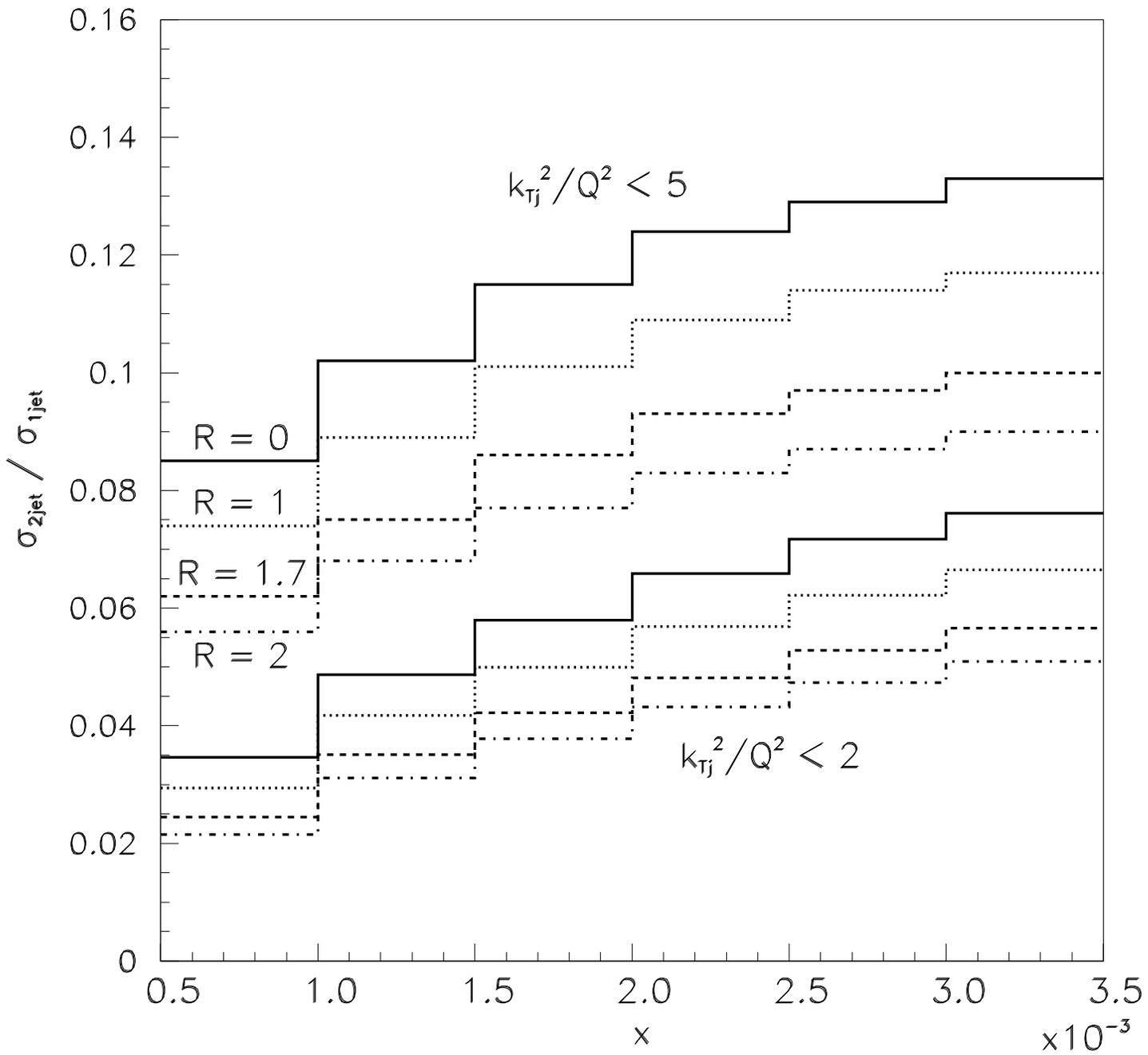
\item[Fig.~6] The same as Fig.~4, but showing the ratio of the
two/one forward jet cross section.

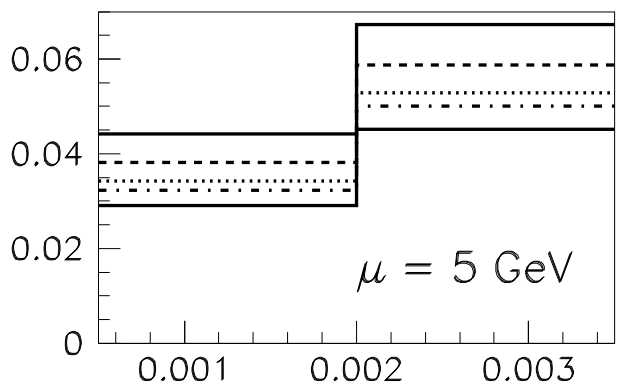
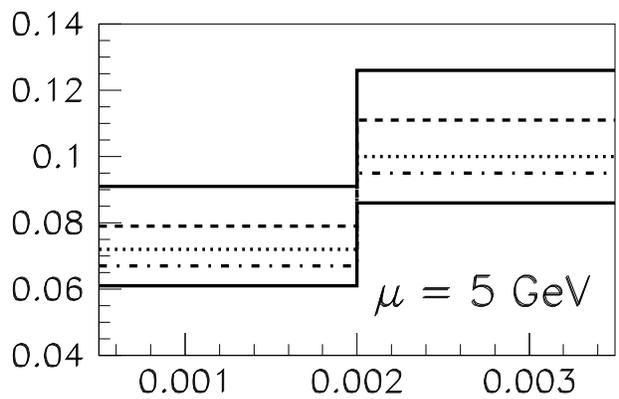
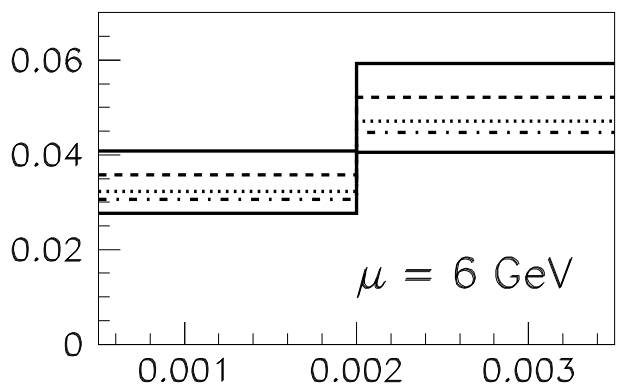
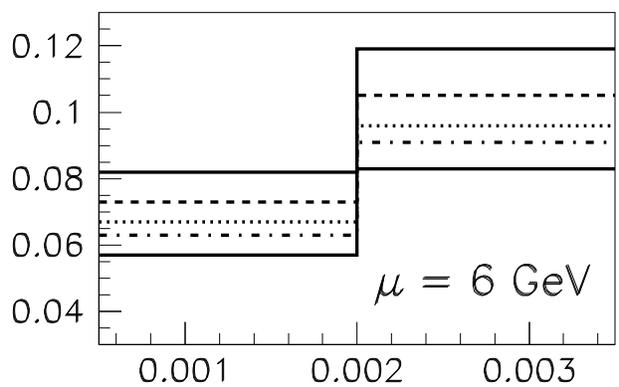
\item[Fig.~7] The two/one forward jet ratio predicted for DIS
events shown for three different values of the jet resolution
parameter $\mu \; ({\rm that \; is} \; k_{Tj}^2 > \mu^2)$, and
for
two
upper
limits of
$k_{Tj}^2 / Q^2$.  The two jets are taken to have the same
$k_{Tj}^2$.  The five different histograms correspond, in
descending order, to $R = 0, \; 1, \; 1.5, \; 1.7$ and 2, where
the jet
separation parameter $R$ is defined in (\ref{eq:a15}).

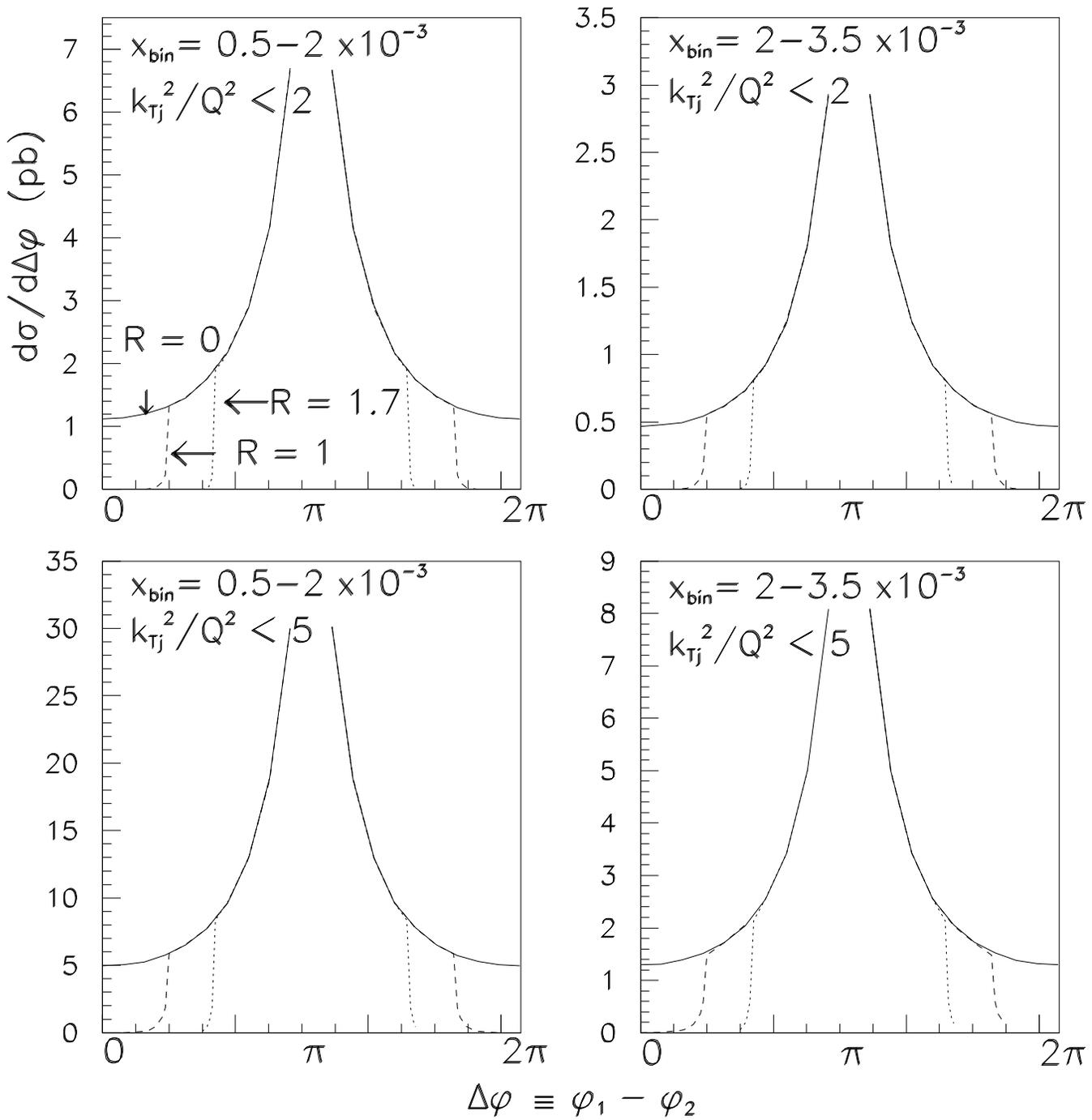
\item[Fig.~8] The cross section as a function of the azimuthal
separation, $\Delta \phi = \phi_1 - \phi_2$, of the two forward
jets.  The distribution is shown for two choices of $k_{Tj}^2 /
Q^2$ and for two different $x$ bins.  We take $y < 0.5$, and we
show the distributions for $R = 0, \: 1$ and 1.7, where the jet
separation parameter $R$ is defined in (\ref{eq:a15}). 
\end{itemize}

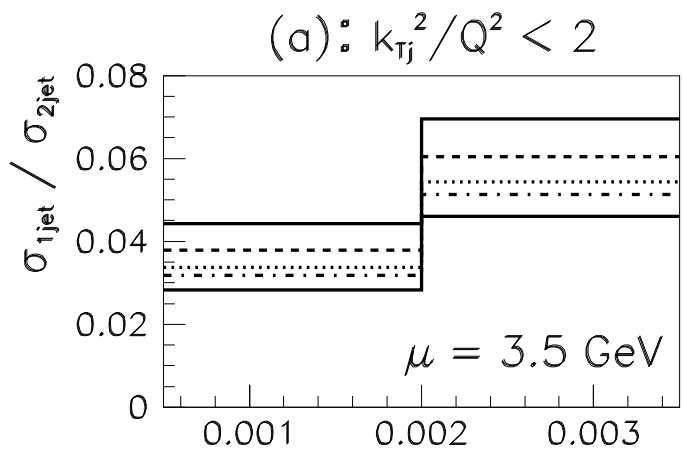
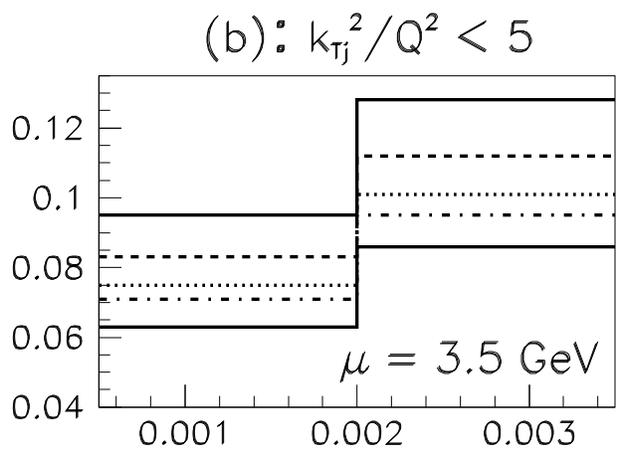
\end{document}